\documentclass[prd,twocolumn,superscriptaddress,nofootinbib,showpacs,amsmath,amssymb]{revtex4}

\usepackage{graphicx,amsmath,amsfonts,amssymb,revsymb,dcolumn,epsfig,bm}

\begin{document}
\title{Energy conditions bounds and their confrontation with supernovae 
data}

\author{M.P. Lima} \email{penna@cbpf.br}
\affiliation{Centro Brasileiro de Pesquisas F\'{\i}sicas, 
Rua Dr.\ Xavier Sigaud 150 \\
22290-180, Rio de Janeiro -- RJ, Brasil}

\author{S. Vitenti}\email{vitenti@cbpf.br}
\affiliation{Centro Brasileiro de Pesquisas F\'{\i}sicas, 
Rua Dr.\ Xavier Sigaud 150 \\
22290-180, Rio de Janeiro -- RJ, Brasil}

\author{M.J. Rebou\c{c}as}\email{reboucas@cbpf.br}
\affiliation{Centro Brasileiro de Pesquisas F\'{\i}sicas, 
Rua Dr.\ Xavier Sigaud 150 \\
22290-180, Rio de Janeiro -- RJ, Brasil}

\date{\today}

\begin{abstract}
The energy conditions play an important role in the understanding
of several properties of the Universe, including the current accelerating
expansion phase and the possible existence of the so-called phantom fields.
We show that the integrated bounds provided by the energy conditions on cosmological
observables such as the distance modulus $\mu(z)$ and the lookback time
$t_L(z)$ are not sufficient (nor necessary) to ensure the local fulfillment
of the energy conditions, making explicit the limitation of these bounds 
in the confrontation with observational data. 
We recast the energy conditions as bounds on the deceleration  and 
normalized Hubble parameters, obtaining new bounds which 
are necessary and sufficient for the local fulfillment of the energy conditions. 
A statistical confrontation, with $1\sigma-3\sigma$ confidence levels,
between our bounds and supernovae data from the \emph{gold} and
\emph{combined} samples is made for the recent past.   
Our analyses indicate, with $3\sigma$ confidence levels, the fulfillment of
both the weak energy condition (\textbf{WEC}) and dominant energy condition 
(\textbf{DEC}) for $z \leq 1$ and $z \lesssim 0.8$, respectively. 
In addition, they suggest a possible recent violation of the null energy condition 
(\textbf{NEC}) with $3\sigma$, i.e. a very recent phase of super-acceleration. 
Our analyses also show the possibility of violation of the strong
energy condition (\textbf{SEC}) with $3\sigma$ in the recent past 
($z \leq 1$), but interestingly 
the $q(z)$-best-fit curve crosses the \textbf{SEC}--fulfillment divider 
at $z \simeq 0.67$, which is a value very close to the beginning of the 
epoch of cosmic acceleration predicted by the standard  concordance 
flat $\Lambda$CDM  scenario. 
\end{abstract}


\maketitle

\section{Introduction}

In classical general relativity, if one wishes to study spacetime properties
that hold for a variety of matter sources it is suitable to impose the 
so-called \emph{energy conditions} that limit the arbitrariness of the 
energy-momentum tensor $T_{\mu\nu}$ on physical grounds. 
These conditions can be stated in a coordinate-invariant way in terms of 
$T_{\mu\nu}$ and vector fields of fixed character (timelike, null and spacelike). 
However, within the framework of the standard Friedmann-Lema\^{\i}tre-Robertson-Walker 
(FLRW) model, we only need to consider 
the energy-momentum tensor of a perfect fluid with density $\rho$ and pressure 
$p\,$, i.e., 
\begin{equation} \label{EM-Tensor}
T_{\mu\nu} = (\rho+p)\,u_\mu u_\nu - p \,g_{\mu \nu}\;,
\end{equation} 
so that the energy conditions take one 
of the following forms~\cite{Hawking-Ellis,Carroll,Wald}:
\begin{equation} \label{ec}  
\begin{array}{lll}
\mbox{NEC}: \  &\, \rho + p \geq 0 \;,  &   \\
\\
\mbox{WEC}: \ \  & \rho \geq 0 &
\ \mbox{and} \quad\, \rho + p \geq 0 \;,  \\
\\
\mbox{SEC}:   & \rho + 3p \geq 0 &
\ \mbox{and} \quad\, \rho + p \geq 0 \;, \\
\\
\mbox{DEC}:    & \rho \geq 0  &
\ \mbox{and} \; -\rho \leq p \leq\rho \;,
\end{array}
\end{equation}
where NEC, WEC, SEC and DEC correspond, respectively, to the null, weak, 
strong and dominant energy conditions. 
Clearly, the  ordinary matter in the form of baryons or relativistic 
particles like photons and neutrinos satisfies these energy conditions. 

{}From the theoretical point of view, the energy conditions have been used 
in different contexts to derive powerful results in a variety of 
situations. For example, the Hawking-Penrose singularity theorems invoke 
the SEC~\cite{Hawking-Ellis}, the positive mass theorem assumes the 
DEC~\cite{Shoen_Yau1981}, while the proof of second law of 
black hole thermodynamics  requires 
NEC~\cite{Visser,Wald}. 

On macroscopic scales relevant for cosmology, another important viewpoint 
is the confrontation of the energy condition predictions with the 
observational data. 
In this regard, since the pioneering works by Visser~\cite{M_Visser1997}, 
it has been shown that  the energy conditions provide model-independent 
bounds on the cosmological observables, and a number of studies involving 
such bounds have been recently discussed in the literature%
~\cite{SAR2006,SAPR2007,Gong-et-al2007,Gong_Wang2007,SARP2007,Sen_Scherrer2007,%
Santiago2006,SARC2007} (see also the related Refs.~\cite{EnergCond_rel}).
Santos \emph{et al.}~\cite{SAR2006,SAPR2007} have derived bounds on the
distance modulus, $\mu(z)\,$,  for any spatial curvature $k$, and made a 
confrontation of the bound predictions with recent type Ia supernovae
(SNe Ia) data. 
In Refs.~\cite{Gong-et-al2007,Gong_Wang2007} the confrontation of 
the NEC and SEC bounds with a \emph{combined} sample of 192 supernovae was 
carried out providing similar and complementary results. They have also
shown that the violation of \emph{integrated} bounds [such as those 
on $\mu(z)\,$] at a given 
redshift $z$ ensures the breakdown of the corresponding energy condition,
without specifying at what redshift the energy-condition violation took place.
In Ref.~\cite{SARP2007} model-independent energy-conditions bounds 
on the lookback time, $t_L(z)\,$,  was derived and
a confrontation with age estimates of galaxies was made.
Sen and Scherrer~\cite{Sen_Scherrer2007} derived upper limits
on the matter density parameter $\Omega_m$ from the WEC in a flat
($k=0)$ universe. 
In the recent Ref.~\cite{CattoenVisser}, Catto\"en and Visser have  
reviewed and complemented some aspects of 
Refs.~\cite{M_Visser1997, SAR2006,SAPR2007,SARP2007}.
Energy conditions constraints on modified gravity models, such as the 
so-called $f(R)$--gravity, have also been investigated in 
Ref.~\cite{Santiago2006} and more recently in Ref.~\cite{SARC2007}. 

In this paper, to proceed further with the investigation of the
interrelation between energy conditions on scales relevant for cosmology 
and observational data, we extend and complement the results of Refs.~\cite{SAR2006,SAPR2007,Gong-et-al2007,Gong_Wang2007,SARP2007} 
in three different ways. 
First, we show in a simple way that the violation 
of  \emph{integrated bounds} such as those on the Hubble
parameter $H(z)$, on  the distance modulus 
$\mu(z)\,$~\cite{SAR2006,SAPR2007} and on the lookback time  
$t_L(z)$~\cite{SARP2007} at a redshift $z$ is neither necessary
nor sufficient \emph{local condition} for the breakdown of the 
associated energy condition~\cite{Gong-et-al2007,Gong_Wang2007}. %
Second, we derive \emph{local}  \emph{necessary and sufficient} 
bounds for the fulfillment of each energy conditions in terms of the deceleration 
parameter $q(z)$ and the normalized Hubble function $E(z)=H(z)/H_0\,$
for any spatial curvature. 
Third, we make the confrontation between our local \emph{non-integrated bounds}  
with statistical estimates  [in the plane $E(z)-q(z)$]
by using the SNe Ia of both the new \emph{gold} sample~\cite{Riess2006}, of 182  SNe Ia  
and the \emph{combined} sample of 192 SNe Ia~\cite{192SNe}. 
In this way, our necessary and sufficient non-integrated energy-condition 
bounds allow a statistical confrontation of energy conditions
and SNe Ia data within chosen confidence levels 
at any given redshift. 

\section{Integrated Bounds from the Energy Conditions}
\label{sec:integrated}

In this section we give an account of our basic assumptions, briefly recast 
the major results of Refs.~\cite{SAR2006,SAPR2007,Gong-et-al2007,Gong_Wang2007},  
and discuss the nature of the energy-condition \emph{integrated} bounds 
and their limitation in the local confrontation with observational data.

Let us begin by recalling that the standard approach to cosmological 
modelling commences with a space-time 
manifold endowed with the Friedmann-Lema\^{\i}tre-Robertson-Walker (FLRW)
metric
\begin{equation}
\label{RWmetric}
ds^2 =  dt^2 - a^2 (t) \left[\, \frac{dr^2}{1-kr^2} + r^2(d\theta^2 
         + \sin^2 \theta  d\phi^2) \,\right],
\end{equation}
where the spatial curvature $k=0,1$ or $-1$ and $a(t)$ is the cosmological 
scale factor. The metric~(\ref{RWmetric}) encodes  the assumption that our 
$3$--dimensional space is homogeneous and isotropic at sufficiently large 
scales along with the existence of a cosmic time $t$. 
However, to study the dynamics of the Universe
an additional assumption in this approach to cosmological modelling is 
necessary, namely, that the large scale structure of the Universe is 
essentially governed by the gravitational interactions, and hence can be 
described by a  metrical theory of gravitation such as  general 
relativity (GR).

These very general premises, which we assume in this work, constrain 
the cosmological fluid to be a perfect-type fluid of the form~(\ref{EM-Tensor}),
with the total density $\rho$ and pressure $p\,$ given by
\begin{eqnarray}
\rho & = & \frac{3}{8\pi G}\left[\,\frac{\dot{a}^2}{a^2}
                                  +\frac{k}{a^2} \,\right]\;,
\label{rho-eq} \\
p & = & - \frac{1}{8\pi G}\left[\, 2\,\frac{\ddot{a}}{a} +
\frac{\dot{a}^2}{a^2} + \frac{k}{a^2} \,\right] \;, \label{p-eq}
\end{eqnarray}
where dots denote derivative with respect to the time $t$. 

The \emph{integrated} bounds on the Hubble functions $H(z)$ 
comes from the following set of dynamical constraints:%
\footnote{In line with the usage in Refs.~\cite{SAR2006,SAPR2007,SARP2007},
here and in what follows we use the boldface-type to denote the energy-condition
restriction that is not contained in any of the previous set of energy-conditions 
inequations [see Eqs.(\ref{ec})]. 
In this way, \textbf{NEC}, \textbf{WEC}, \textbf{SEC} and \textbf{DEC} 
refer, respectively, to the following NEC, WEC, SEC and DEC inequations: 
$\rho + p \geq 0\,$, $\rho \geq 0\,$, $\rho + 3p \geq 0\,$, and 
$\rho - p \geq 0\,$.} 
%
\begin{eqnarray}
\mbox{\bf NEC} & \, \Rightarrow &  - \frac{\ddot{a}}{a}
+  \frac{\dot{a}^2}{a^2}  + \frac{k}{a^2} \geq 0 \;, 
\label{nec-eq} \\ 
\mbox{\bf WEC} & \, \Rightarrow &  \frac{\dot{a}^2}{a^ 2} 
+ \frac{k}{a^2} \geq 0 \;,
\label{wec-eq} \\
\mbox{\bf SEC} & \, \Rightarrow &  \frac{\ddot{a}}{a} \leq 0 \;,
\label{sec-eq} \\
\mbox{\bf DEC} & \, \Rightarrow &  \frac{\ddot{a}}{a} + 2\left[
\frac{\dot{a}^2}{a^2}+\frac{k}{a^2} \right] \geq 0 \;, \label{dec-eq}
\end{eqnarray}
which can be easily derived from the energy conditions [Eqs.~\eqref{ec}] along 
with the above Eqs.~\eqref{rho-eq} and \eqref{p-eq}.
In fact, Eqs.~\eqref{nec-eq}--\eqref{dec-eq} can be written in terms of the 
Hubble function, $H(z) = \dot{a}(t)/a(t)$, and its derivatives with respect 
to the redshift, $z =(a_0/a) - 1$, as 
\begin{eqnarray}
\label{eq:nec}
\mbox{\bf NEC} & \, \Rightarrow & \quad\, \frac{\partial H^2}{\partial z} 
 \geq - \frac{2k(1 + z)}{a_0^2}\;, \\
\label{eq:wec}
\mbox{\bf WEC} & \, \Rightarrow & \quad\, -\frac{k(1+z)^2}{a_0^2H^2} 
\leq 1 \;, \\
\label{eq:sec}
\mbox{\bf SEC} & \, \Rightarrow & \quad\, \frac{\partial \log H^2}{\partial z} 
\geq \frac{2}{(1 + z)}\;, \\
\label{eq:dec}
\mbox{\bf DEC} & \, \Rightarrow & \quad\, \frac{\partial}{\partial z} 
\left(\frac{H^2}{(1 + z)^6}\right) \leq \frac{4k}{a_0^2(1 + z)^5}\;,
\end{eqnarray}
where here and in what follows the subscript $0$ stands for present-day 
quantities.
Now, integrating the inequations~\eqref{eq:nec}, \eqref{eq:sec} and%
~\eqref{eq:dec} in the interval $(0, z)$, where we assume that they hold, 
one obtains the following  \emph{integrated} bounds
on Hubble function from the energy conditions:
\begin{eqnarray} \vspace{-3mm}
\label{eq:hnec}
\mbox{\bf NEC}  \,& \Rightarrow & H(z) \geq H_0 \sqrt{1 - \Omega_{k0} + 
\Omega_{k0} (1 + z)^2}\;, \\
\nonumber \\
\label{eq:hsec}
\mbox{\bf SEC}  \,& \Rightarrow  & H(z) \geq H_0 (1 + z) \;,  \\
& & \nonumber \\
\label{eq:hdec}
\mbox{\bf DEC} & \Rightarrow & H(z) \leq H_0 (1 + z) \sqrt{(1 - \Omega_{k0})(1 + z)^4 
+ \Omega_{k0}]}\,. \nonumber  \\ 
\end{eqnarray} 

\vspace{-2mm} \noindent
We note that the inequation~\eqref{eq:wec} does not contain 
derivative of $H(z)$, but clearly for $z = 0$ the \textbf{WEC} 
restricts the present-day curvature parameter  to 
$\Omega_{k0} \equiv -k/(a_0H_0)^2 \leq 1$.  

The \emph{integrated} bounds provided by the energy conditions on the
distance modulus $\mu(z)$ can now be easily obtained from the above
bounds on the Hubble function as follows.
First, we recall that the distance modulus for an object at redshift 
$z$ is defined by
\begin{equation} \label{eq:mu_z}
\mu(z) \equiv m(z) - M = 5\, \log_{10}\left( \frac{d_L(z)}{1\text{Mpc}} \right) 
+ 25\;,          
\end{equation}
where $m$ and $M$ are, respectively, the apparent and absolute 
magnitudes, and $d_L$ is given by 
\begin{equation}
\label{eq:dl_z}d_L(z) = \frac{c}{H_0}\frac{(1+z)}{\sqrt{\mid\Omega_{k0}\mid}}\;
\mathrm{S}_k\left(\sqrt{\mid\Omega_{k0}\mid}\int_0^z \frac{\mathrm{d}z^\prime}
{E(z^\prime)}\right), 
\end{equation}
where $\mathrm{S}_k(x) = \sin(x), \,x,\, \sinh(x)$ for $k = 1, 0, -1$ respectively,
and $E(z)$ = $H(z)/H_0\,$. 
Second, we substitute Eqs.~\eqref{eq:hnec}--\eqref{eq:hdec} into 
Eqs.~(\ref{eq:dl_z}) and  (\ref{eq:mu_z}) to obtain the bounds
on the distance modulus $\mu(z)$ for any spatial curvature $k$. 
For the flat FLRW model ($\Omega_{k0} = 0$) which we focus our attention 
on in this paper, the \emph{integrated} bounds reduce to
\begin{eqnarray}
\label{eq:munec}
\mbox{\bf NEC} & \Rightarrow & \mu(z) \leq 5 \log_{10} [cH_0^{-1} z 
(1 + z)] + 25\;,\\
\nonumber \\
\label{eq:musec}
\mbox{\bf SEC}  & \Rightarrow & \mu(z) \leq 5 \log_{10} [cH_0^{-1} (1 + z) 
\ln(1 + z)] + 25 \;, \nonumber \\
\\
\label{eq:mudec}
\mbox{\bf DEC} & \Rightarrow & \mu(z) \geq 5 \log_{10} \left[ \frac{cH_0^{-1} z 
(2 + z)}{2 (1 + z)} \right] + 25 \;.
\end{eqnarray}

Concerning the above bounds on $H(z)$ and $\mu(z)\,$, 
we emphasize that the \emph{nonlocal} or \emph{integrated} nature  
of these bounds arises from the fact that they were obtained by assuming 
the fulfillment of the energy condition in the whole interval 
of integration  $(0,z)\,$. However, in the same way that a positive sum of $N$ 
terms does not necessarily imply that all the terms of the sum are also 
positive, the fulfillment of the \emph{integrated} bounds on $H(z)$ and $\mu(z)$ 
does not necessarily imply that the energy conditions are obeyed in 
all subintervals of $(0, z)$ but only in at least an undetermined subinterval. 
Reciprocally, the violation of these \emph{integrated} bounds 
merely implies that the corresponding energy condition was violated in at least a 
subinterval of $(0, z)$.   
This amounts to saying that the fulfillment (or the violation) of any of 
these bounds at a given redshift $z$ is not a sufficient (nor a necessary)
\emph{local} condition for the fulfillment (or respectively the violation) 
of the associated energy condition at $z$. In practice, this means that 
the local confrontation between the prediction of the \emph{integrated} bounds 
such as those on $H(z)$  and on $\mu(z)$  [Eqs.~\eqref{eq:hnec}--\eqref{eq:hdec}] 
and [{Eqs.~\eqref{eq:munec}--\eqref{eq:mudec}] and observational 
data  is not suitable to draw conclusions on the local fulfillment 
or violation of the associated energy conditions at $z$.%
%
\footnote{We note that at a \emph{nonlocal} level, the fulfillment 
(or the violation) of each of these \emph{integrated} bounds at a given $z$ is 
sufficient to ensure only the fulfillment (or respectively the violation) 
of the associated energy conditions somewhere in at least a subinterval of the
integration interval $(0,z)\,$, 
as discussed in Refs.~\cite{Gong-et-al2007,Gong_Wang2007} and concretely
illustrate in Section~\ref{sec:results}. }

\section{Non-integrated Bounds from the energy conditions}
\label{sec:new_bounds}

The practical limitation in the local confrontation between the above 
\emph{integrated} bounds and observational data calls for \emph{non-integrated} 
bounds from energy conditions, which can be easily obtained by 
rewriting Eqs.~\eqref{nec-eq}--\eqref{dec-eq} in terms of the deceleration 
parameter, $q(z) = -\ddot{a}/aH^2$, and the normalized Hubble function, 
$E(z) = H(z)/H_0$, in the following form: 
\begin{eqnarray}
\label{eq:nec-q(z)}
\mbox{\bf NEC} & \,\Leftrightarrow & \;\, q(z) - \Omega_{k0} 
\frac{(1+z)^2}{E^2(z)}   \,\geq -1 \;, \\
\label{eq:wec-omega} 
\mbox{\bf WEC} & \, \Leftrightarrow & \;\, \frac{E^2(z)}{(1 + z)^2} 
\,\geq \Omega_{k0} \;, \\
\label{eq:sec-q(z)} 
\mbox{\bf SEC} & \, \Leftrightarrow & \;\, q(z) \,\geq 0 \;, \\
\label{eq:dec-q(z)}  
\mbox{\bf DEC} & \, \Leftrightarrow & \;\, q(z) + 2\,\Omega_{k0}
\,\frac{(1+z)^2 }{E^2(z)} \,\leq  2 \;,
\end{eqnarray}
for any  spatial curvature $\Omega_{k0}\,$.

Some words of clarification are in order here concerning the above bounds. 
First, we note that for a fixed value of $\Omega_{k0}$, Eq.~\eqref{eq:wec-omega} 
provides the \textbf{WEC} lower-bound on normalized Hubble function $E(z)$
for any $z$, whereas Eqs.~\eqref{eq:nec-q(z)} and \eqref{eq:dec-q(z)} give, 
respectively, the \textbf{NEC} and \textbf{DEC} bounds on parameters of  
$E(z)-q(z)$ plane for any fixed redshift $z_\star$ (say). Also, the \textbf{SEC} 
lower bound [Eq.~(\ref{eq:sec-q(z)})] clearly holds regardless of 
the value of the spatial curvature. Second, since the bounds have been derived 
without making any integration (\emph{non-integrated} bounds), they 
are clearly sufficient and necessary to ensure the \emph{local} fulfillment 
of the associated energy condition.  
In practice, this allows local confrontation between the predictions 
of these \emph{non-integrated} bounds and, e.g.,  SNe Ia data, an issue which we 
shall discuss in the following sections focusing in the flat ($k=0$) FLRW  case, 
in which the \textbf{NEC}  and \textbf{DEC} \emph{non-integrated} bounds  
reduce, respectively, to $q(z) \geq -1\,$ and $q(z) \leq 2\,$, and where
obviously the fulfillment of the \textbf{SEC} [$\,q(z) \geq 0\,$] implies 
that the \textbf{NEC} is satisfied identically. 

\section{Analysis and Discussion}
\label{sec:analysis}

\begin{figure*}[ht]
\includegraphics[scale=0.8]{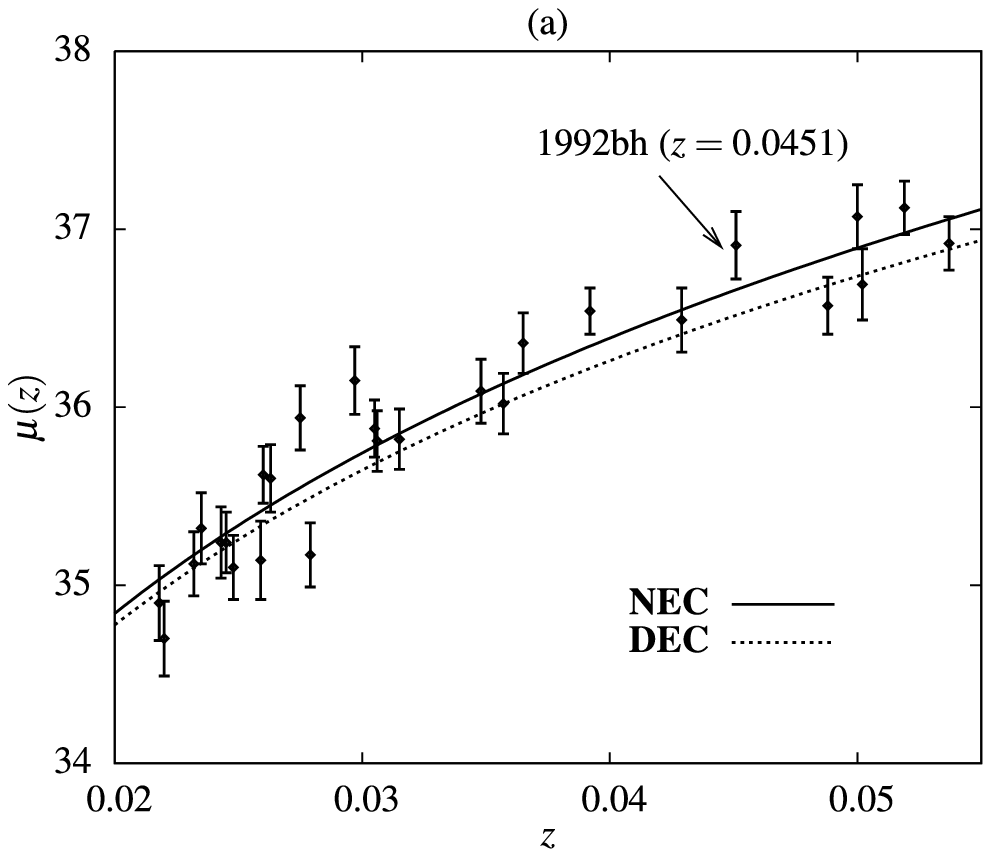}
\includegraphics[scale=0.8]{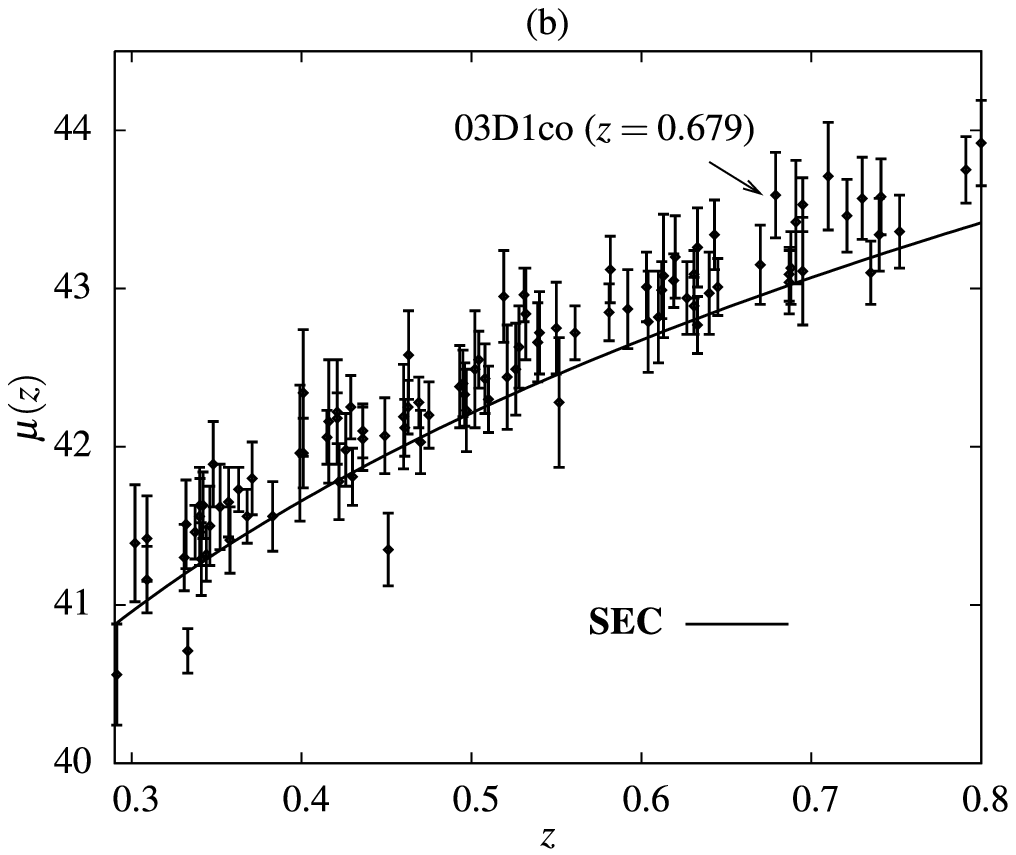}
\caption{Panel (a): The \textbf{NEC} upper-bound and the \textbf{DEC} lower-bound on the 
distance modulus, $\mu(z)$, in the redshift interval $(0.02, 0.055)$. Panel (b): 
The \textbf{SEC} upper-bound on $\mu(z)$ in the  redshift interval $(0.3, 0.8)$.
The data points in both panels correspond to type Ia supernovae from a \emph{combined} 
sample, and consistently we have taken $H_0 = 65.8 \, \text{km} \, \text{s}^{-1} \, \text{Mpc}^{-1}$.}
\label{fig:modulus}
\end{figure*}

\begin{figure*}[ht]
\includegraphics[scale=0.8]{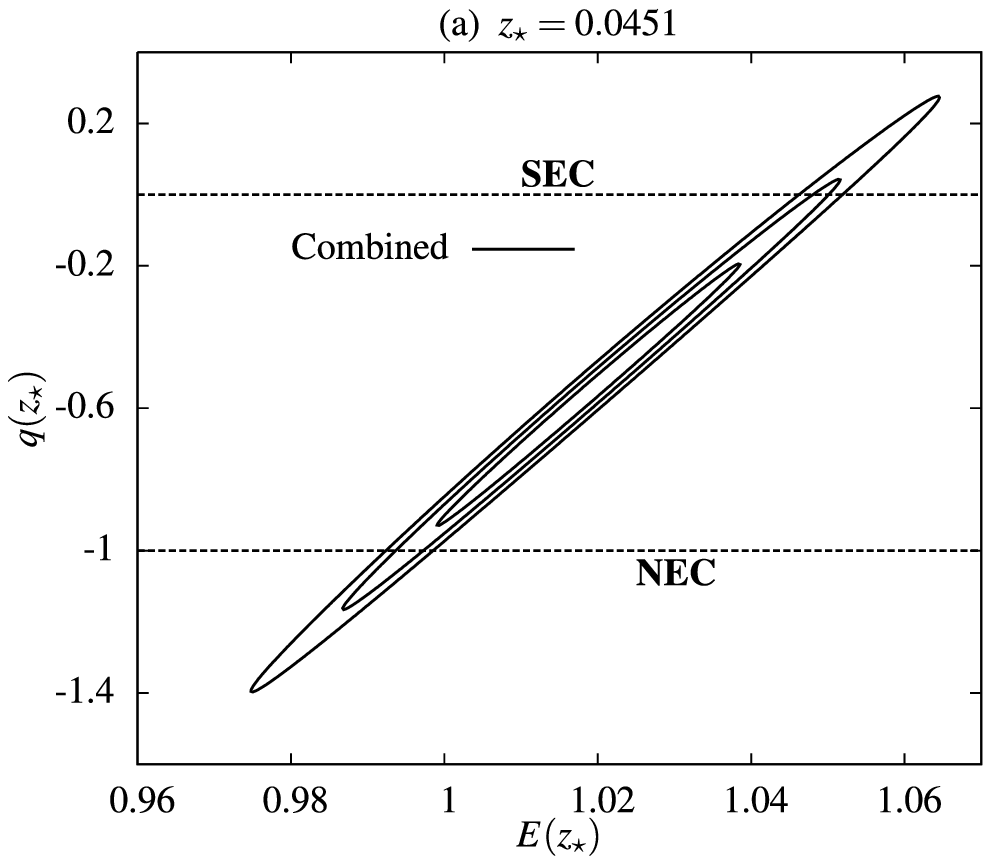}
\includegraphics[scale=0.8]{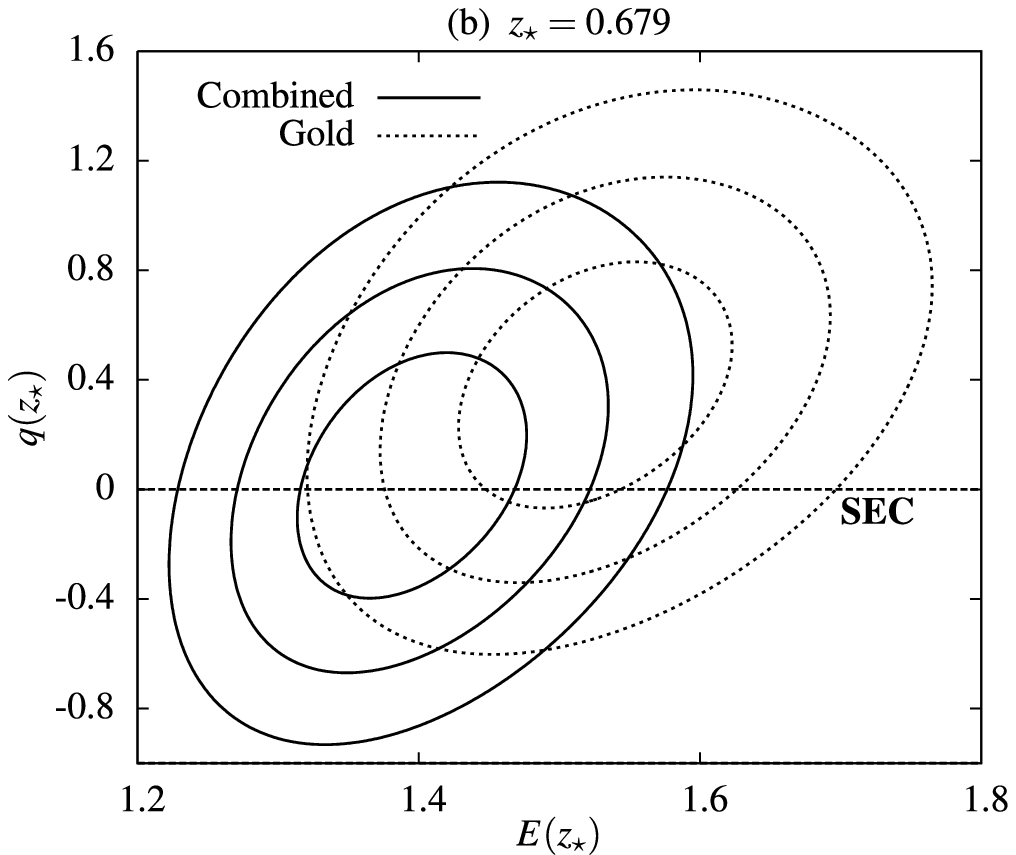}
\caption{The $1\sigma$, $2\sigma$ and $3\sigma$ contours obtained with the $E(z_\star)$ 
and $q(z_\star)$ estimates from \emph{combined} (solid lines) and \emph{gold} (dashed lines) 
samples at $z_\star = 0.0451$ [panel (a)] and at $z_\star = 0.679$ [panel (b)].
The \emph{non-integrated} \textbf{SEC} and \textbf{NEC} bounds are also indicated. 
The best fit values for $(\,E(z),q(z)\,)$ at $z_\star = 0.0451$ are $(1.015, -0.638)$ 
for \emph{gold}, and $(1.020, -0.559)$ for \emph{combined}. At $z_\star = 0.679$
the best fit values are $(1.621, 0.469)$ for \emph{gold}, and $(1.464, 0.215)$ for 
\emph{combined} sample.}
\label{fig:bounds_2sigma}
\end{figure*}

\subsection{Estimates of $q(z_\star)$ and $E(z_\star)$}
\label{sec:data-analysis}

In order to make the confrontation between the \emph{non-integrated} bounds
[Eqs.~\eqref{eq:nec-q(z)}--\eqref{eq:dec-q(z)}] with observational data
we need estimates of $q(z_\star)$ and $E(z_\star)$. 
Model-independent estimates of these parameters can be obtained by
approximating the deceleration parameter as a function of the redshift in 
terms of a linear piecewise continuous function, known as linear spline,
\begin{equation}
\label{eq:q_z}
q(z) = q_l + q^\prime_l \Delta z_l \;, \quad z \in (z_l, z_{l+1}),
\end{equation}
where the subscript $l$ means that the quantity is taken at $z_l\,$, $\Delta z_l 
\equiv (z-z_l)\,$, and the prime means the derivative with respect to $z$. 
We use the definition of $q(z)$ in terms of $H(z)$ to obtain
\begin{equation}
\label{eq:E_z}
E(z) = \exp{\int_0^z \frac{1+q(z)}{1+z}\mathrm{d}z},
\end{equation}
and, consequently, the luminosity distance and the distance modulus using
Eqs.~\eqref{eq:dl_z} and \eqref{eq:mu_z}. Then we fitted the parameters
of the $q(z)$ curve using the type Ia supernovae (SNe Ia) redshift--distance 
modulus data from \emph{gold} sample \cite{Riess2006} and a \emph{combined} sample 
\cite{192SNe}.

\subsection{Results}
\label{sec:results}

In Figs.\ref{fig:modulus}(a) and~\ref{fig:modulus}(b) we confront,
the \textbf{NEC} along with  \textbf{DEC}, and the \textbf{SEC} 
\emph{integrated} bounds on $\mu(z)$ [Eqs.~(\ref{eq:munec}), (\ref{eq:mudec}) and~(\ref{eq:musec})] 
with SNe Ia of the \emph{combined} sample 
as compiled in Ref.~\cite{192SNe} for, respectively, the redshift intervals 
$(0.02, 0.055)$ and $(0.3, 0.8)$ by taking consistently
$H_0 = 65.8 \,\,\text{km}\,\,\text{s}^{-1}\,\text{Mpc}^{\!-\!1}\,$.%
\footnote{We note that for the \emph{combined} sample provided by Riess at \url{http://braeburn.pha.jhu.edu/~ariess/R06/}, the distance modulus is 
computed using an arbitrary choice of the absolute magnitude $M$ as 
discussed in \cite{Riess2006,192SNe}.  In Fig.\ref{fig:modulus}(a) 
and~Fig.\ref{fig:modulus}(b) we have dealt with this arbitrariness 
by noting that the confrontation between the integrated bounds 
[Eqs.~(\ref{eq:munec})--(\ref{eq:mudec})] with SNe Ia data depends 
on $H_0$ and $M$ through the additive term $m_0 = M+5\log_{10}(\frac{c/H_0}{1\rm{Mpc}})$.
Thus, to obtain a value of $m_0$ consistent with the SNe data, we have fitted 
the low redshift ($z \leq 0.3$)  SNe Ia distance modulus 
treating $H_0$ as an unknown and taking the values of magnitude $M$ as given  
by the \emph{combined} sample. 
Clearly, this procedure for comparison between bounds with SNe Ia data
is independent of a particular value of $H_0$ in the sense that  
one can also begin by taking a specific value of $H_0$, adjust 
the value of $m_0$, and treating $M$ as unknown instead.}

Clearly there are several SNe data points indicating the violation
with more than $1\sigma$ of the \emph{integrated} \textbf{NEC} and \textbf{SEC} 
upper-bounds on $\mu(z)\,$, and two data points suggesting the violation 
of the \emph{integrated} \textbf{DEC} lower-bound on the distance modulus with 
$\gtrsim 1\sigma$. 

To concretely contrast the \emph{integrated} and \emph{non-integrated} bounds by 
using supernovae observations, we consider, as a first example, the 
supernova 1992bh at $z=0.0451$ of the \emph{combined} sample, %
whose observed distance modulus is $\mu_{\rm 1992bh}= 36.91 \pm 0.19$
while the upper-bound \textbf{NEC} predictions is $\mu(z=0.0451)= 36.66$, i.e.,
it violates  the integrated upper-bound from \textbf{NEC} 
with  $\simeq 1.31\sigma\,$ [see Fig.\ref{fig:modulus}(a)]. 
However, 
this local violation of the \textbf{NEC} \emph{integrated} bounds is
not sufficient to guarantee the breakdown of the \textbf{NEC} at $z = 0.0451$.
This point is made apparent in Fig.~\ref{fig:bounds_2sigma}(a) which 
shows the $1\sigma$, $2\sigma$ and $3\sigma$ confidence regions in the plane 
$E(z)-q(z)$ [estimated from \emph{combined} sample at $z_\star = 0.0451\,$]
along with the non-integrated (local) \textbf{NEC} bound [$\,q(z) \geq -1\,$]. 
Clearly, the fact that the whole $1\sigma$ confidence region is above 
the \textbf{NEC} \emph{non-integrated} bound is sufficient to ensure the
fulfillment of the \textbf{NEC} with $1\sigma$  at $z_\star = 0.0451$,
despite the $\simeq 1.31\sigma\,$ violation of the \textbf{NEC} \emph{integrated} 
bound at this redshift. 

Figures~\ref{fig:modulus}(b) and \ref{fig:bounds_2sigma}(b) show
the contrast between the \textbf{SEC} \emph{integrated} and \emph{non-integrated} 
bounds. In Fig.~\ref{fig:modulus}(b) the supernova 03D1co  
(which belongs to the \emph{combined} and \emph{gold} samples) at $z_\star = 0.679$, 
is such that its observed value of the distance modulus
[\emph{gold}: $\mu = 43.58 \pm 0.19\,$, \emph{combined}: $\mu = 43.59 \pm 0.27\,$] 
violated the \textbf{SEC} \emph{integrated} upper-bound predictions
[\emph{gold}: $\mu = 43.06\,$, \emph{combined}: $\mu = 42.99\,$] with $2.22\sigma$ 
(\emph{gold}) and $2.72\sigma$ (\emph{combined}). Fig.~\ref{fig:bounds_2sigma}(b)
shows the $1\sigma$, $2\sigma$ and $3\sigma$  confidence regions in the plane 
$E(z)-q(z)$ [estimated from \emph{gold} and \emph{combined} sample at $z_\star = 0.679\,$]
along with the \emph{non-integrated} \textbf{SEC} lower-bound [$\,q(z) \geq 0\,$].
The comparison of these figures makes clear that
although the observed $\mu(z)$ values are more than $2\sigma$
higher than the \textbf{SEC} \emph{integrated} upper-bounds
(suggesting at first sight a violation of the \textbf{SEC}) 
the \emph{non-integrated} bound analysis shows that  at $z_\star = 0.679\,$  
the \textbf{SEC} can either be fulfilled or violated within $1\sigma$,  
$2\sigma$ and $3\sigma$ confidence levels for both SNe Ia samples.

In order to obtain a detailed global picture of the breakdown and 
fulfillment of the energy conditions in the recent past, we shall extend  
the above local analysis by examining the behavior of the \emph{non-integrated} 
energy-condition bounds with $1\sigma - 3\sigma$  confidence levels for the recent 
past ($0< z \leq 1$) using the \emph{combined} and \emph{gold} SNe Ia samples.
To this end, we first divide the redshift interval $(0,1]$ into $100$ equally 
spaced points at which we carry out the statistical estimates and 
confrontation of the \emph{non-integrated} bounds with SNe Ia data. 
Second, we note that, for the flat case, the \textbf{NEC}, \textbf{SEC} and 
\textbf{DEC} \emph{non-integrated} bounds do not dependent on the estimates
of $E(z_\star)$ [see Eqs.~\eqref{eq:nec-q(z)}, \eqref{eq:sec-q(z)}, \eqref{eq:dec-q(z)}
and Fig.~\ref{fig:bounds_2sigma}], and therefore the upper and lower 
$1\sigma - 3\sigma$ limits of $q(z_\star)$ are sufficient to establish the 
fulfillment or violation of these energy conditions within 
these confidence levels.  

\begin{figure*}[ht]
\includegraphics[scale=0.8]{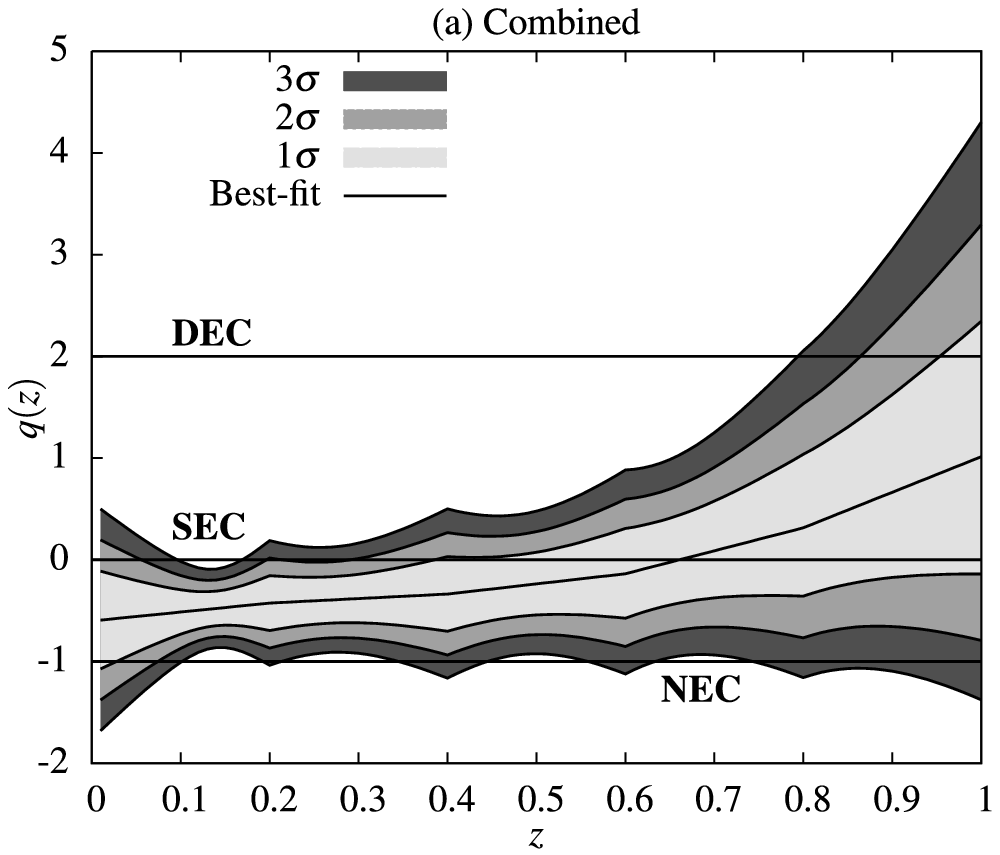}
\includegraphics[scale=0.8]{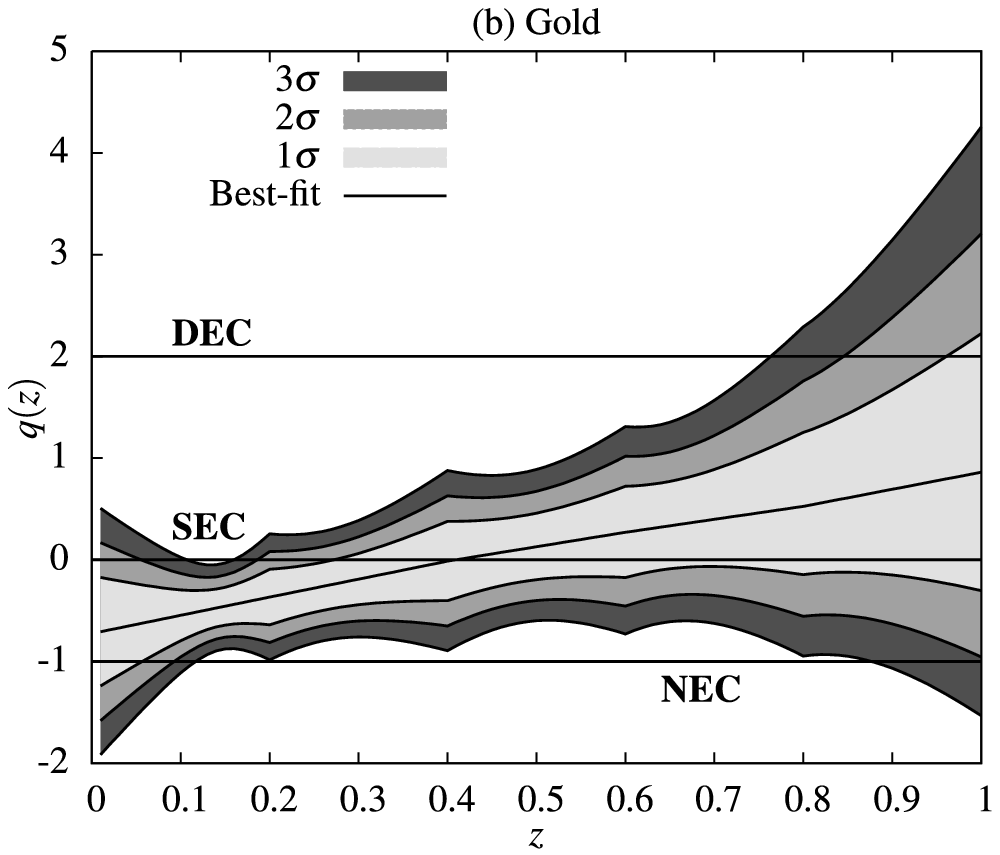}
\caption{The best-fit, the upper and lower $1\sigma$, $2\sigma$ and $3\sigma$
limits of $q(z)$ for $100$ equally spaced redshifts. The \textbf{NEC} and \textbf{SEC} 
\emph{non-integrated} lower-bounds, and also the \textbf{DEC} 
\emph{non-integrated} upper-bound for the flat case are shown. 
This figure shows that the \textbf{SEC} is violated with $1\sigma$ confidence 
level until $z \simeq 0.38$ for \emph{combined} [panel (a)] and until 
$z \simeq 0.26$ for \emph{gold} sample [panel (b)]. It shows 
the violation of the  \textbf{SEC} with $3\sigma$ for low redshift intervals.
It also shows that the \textbf{NEC} and the \textbf{DEC} are violated 
within $3\sigma$ confidence level for, respectively, very low and high 
redshifts. See the text for more details.}
\label{fig:evolution}
\end{figure*}

The two panels in Fig.~\ref{fig:evolution} show the best-fit values and 
$1\sigma$, $2\sigma$ and $3\sigma$ limits for $q(z_\star)$ for the 
\emph{combined} [panel (a)] and \emph{gold} [panel (b)] samples 
along with the \emph{non-integrated} \textbf{NEC}, \textbf{SEC} and 
\textbf{DEC} bounds in the plane $q(z) - z$. 
These panels indicate the violation of the \textbf{SEC} 
with more than $3\sigma$ confidence level in the redshift intervals 
$(\simeq 0.09,\simeq 0.17)$ and $(\simeq 0.11, \simeq 0.16)$ for, 
respectively, the \emph{gold} and \emph{combined} samples. We note
that highest evidence for the violation of \textbf{SEC} is at $z \sim 0.135$ 
for both samples [$\,3.86\sigma$ (\emph{combined}) and $3.43\sigma$ (\emph{gold}) below
the bound$\,$]. 
Clearly, violation of the \textbf{SEC} is also permitted (within $1\sigma$ to
$3\sigma$) for higher redshifts, but the  best-fit $q(z_\star)$  
curves cross the \textbf{SEC}--fulfillment divider at $z \simeq 0.67$ and
$z \simeq 0.42$ for the \emph{combined} and \emph{gold} samples, respectively.

Concerning the \textbf{NEC}, the panels of Fig.~\ref{fig:evolution} show 
its violation within $3\sigma$ for low redshifts [$\,z \in (0, \simeq 0.1)\,$]
for combined and gold samples. 
For higher values of redshift we have the \textbf{NEC}-fullfilment with 
$2\sigma$ for both samples.%
\footnote{Since the violation of the \emph{integrated} bounds at any 
$z$ ensures the violation of the associated energy condition 
in a subinterval $(0,z)$,  the violation of the \textbf{NEC}
within $1\sigma$, in $z \in (0,0.02)$ is the cause for the violation of 
the \textbf{NEC} \emph{integrated} bound by the supernova 1992bh at 
$z =0.0451\,$  of the \emph{combined} sample.
However, differently from the \textbf{NEC} case which is fulfilled with 
$1\sigma$ for $z > 0.05$, Fig.~\ref{fig:evolution} shows no redshift where 
\textbf{SEC} is obeyed with $1\sigma$. In this way, due to the
degeneracies of the SNe Ia data, 
one cannot specify a 
subinterval of $(0,0.679)$ responsible for the violation of the 
\textbf{SEC} \emph{integrated} bound by SNe Ia 03D1co at 
$z=0.679$.}
Regarding the \textbf{DEC},  Fig.~\ref{fig:evolution}  shows that 
it is fulfilled in nearly the whole redshift interval for both samples, 
but it might be violated within 
$3\sigma$ for high redshifts ($z \gtrsim 0.8$)},
where the errors in our estimates grow significantly, though. 

Concerning the above analyses it is worth emphasizing that they are very 
insensitive to the values of the curvature parameter, i.e.,  all the 
above conclusions remain essentially unchanged for values of $\Omega_{k0}$ 
lying in the interval provided by the WMAP and other 
experiments~\cite{Spergel:2006hy}.
In other words, our estimates of $E(z_\star)$, $\,q(z_\star)$ and of the 
\emph{non-integrated} bounds [Eqs.~\eqref{eq:nec-q(z)}--\eqref{eq:dec-q(z)}] 
by using the best-fit value, the upper or the lower $1\sigma$ limits for 
$\Omega_{k0} = -0.014 \pm {0.017}$~\cite{Spergel:2006hy} are very close to 
estimates of those parameters in the flat case, with differences much 
smaller than the associated errors. 

Finally, we also note that  the \emph{non-integrated} \textbf{WEC} 
bound [Eq.~\eqref{eq:wec-omega}] is fulfilled in the whole redshift 
interval $(0,1)$ for the upper $1\sigma$ limit value of the curvature
provided by WMAP team, i.e., $E^2(z) \geq 0.003\,(1 + z)^2$ holds 
for all our estimated values of $E(z_\star)$, whereas for 
the $\Omega_{k0} \in (-0.031,0)$ the \textbf{WEC} is 
fulfilled identically, i.e., regardless of the values of $E(z_\star)$ 
and $z_\star$.

\section{Concluding Remarks} 

By using the fact the classical energy conditions can be recast as a set 
of differential constraints involving the scale factor $a(t)$ and its derivatives
[see (\ref{nec-eq})--(\ref{dec-eq})],  model-independent \emph{integrated} bounds 
on, e.g., the Hubble parameter $H(z)$, the distance modulus $\mu(z)$, and on the 
lookback time  $t_L(z)$ have been recently derived and confronted with 
observational data (see, e.g., Refs.~\cite{M_Visser1997,SAR2006,SAPR2007,%
Gong-et-al2007,Gong_Wang2007,SARP2007,CattoenVisser}). 

In this paper, we have shown that the violation (or the fulfillment)
of these \emph{integrated} bounds at a given redshift $z$ is neither sufficient
nor necessary to ensure the violation (or respectively the fulfillment) of the
energy conditions at $z$.  
In practice, this means that the local confrontation 
between the prediction of the \emph{integrated} bounds and observational data
(such as, e.g.,  those in Refs.~\cite{SAR2006,SAPR2007,SARP2007}) 
is not sufficient to draw conclusions on the violation or fulfillment 
the energy conditions at $z$. This feature is also made apparent in 
Figs.~\ref{fig:modulus} and \ref{fig:bounds_2sigma}, where we present 
concrete examples of violation of \emph{integrated} bounds with either 
fulfillment of the \emph{non-integrated} bounds with $1\sigma$
[panels~\ref{fig:modulus}(a) and \ref{fig:bounds_2sigma}(a)], or
fulfillment and violation of the \emph{non-integrated} bounds with 
$1\sigma$, $2\sigma$  and $3\sigma$ [panels~\ref{fig:modulus}(b) and 
\ref{fig:bounds_2sigma}(b)].

To overcome the crucial drawback in the confrontation between
\emph{integrated} bounds on cosmological observables and observational
data, we have formulated new bounds from energy conditions in terms
of the normalized Hubble and deceleration parameters [$E(z)$ and $q(z)$]
which are necessary and  sufficient for the fulfillment of the
energy conditions [Eqs.~(\ref{eq:nec-q(z)})~--~(\ref{eq:dec-q(z)})]. 
We have also confronted our \emph{non-integrated} bounds with 
model independent estimates of $q(z)$ and $E(z)$ which were obtained by using 
the \emph{gold} sample of 182 SNe Ia provided by Riess \emph{et al.\/} in Ref.~\cite{Riess2006} and with a \emph{combined} sample of 192 SNe Ia 
provided by Wood-Vasey \emph{et al.}~\cite{192SNe} [Figs.~\ref{fig:bounds_2sigma}
and \ref{fig:evolution}]. On general grounds, our analyses indicate the
\textbf{WEC} fulfillment in the recent past ($z \leq 1$) with $3\sigma$,
and a possible recent phase of super-acceleration (violation of the \textbf{NEC}
with $3\sigma$ for $z \in (0,0.1)$ for both the \emph{combined} and  
\emph{gold} samples. 
Our analyses also show that  the \textbf{DEC} is fulfilled  
with $3\sigma$ for all recent past redshifts but $z \geq 0.8\,$.
Concerning the \textbf{SEC} our analyses indicate the possibility
its violation with $1\sigma - 3\sigma$ confidence levels for $z \leq 1$,
with small subintervals in which there is no \textbf{SEC}-fulfillment   
with $3\sigma$ for both the \emph{combined} and  \emph{gold} samples.
An interesting fact from the confrontation between the \textbf{SEC} 
\emph{non-integrated} bound and  SNe Ia \emph{combined} sample is that 
although the violation of the \textbf{SEC} is permitted in the recent 
past  with  $3\sigma$ confidence level, the estimated $q(z)$-best-fit 
curve crosses the \textbf{SEC}--fulfillment divider at $z \simeq 0.67$ 
[see panel~\ref{fig:evolution}(a)], which is very close to redshift of
the beginning of the epoch of cosmic acceleration predicted by the current 
standard  concordance flat $\Lambda$CDM  scenario with $\Omega_m \simeq 0.3$. 

Finally, we emphasized that although we have focused  our attention on the 
flat FLRW case, the above results concerning the new 
\emph{non-integrated} bounds analyses remain unchanged for values 
of $\Omega_{k0}$ lying in the interval provided by  WMAP 
team~\cite{Spergel:2006hy}.

\begin{acknowledgments}
This work is supported by Conselho Nacional de Desenvolvimento Cient\'{i}fico e 
Tecnol\'ogico (CNPq) -- Brasil, under grant No.\ 472436/2007-4. 
M.P.L., S.V. and M.J.R. thank CNPq for the grants under 
which this work was carried out.
M.J.R. also thanks J. Santos for valuable comments.  
\end{acknowledgments}


\end{document}